\documentclass[aps,pra,twocolumn,reprint,letterpaper,superscriptaddress,showpacs]{revtex4} 
\usepackage{amsmath}
\usepackage{amsfonts}
\usepackage{amssymb}
\usepackage{epsfig}
\usepackage{graphicx}

\newcommand{\eff}[0]{\ensuremath{\mathrm{eff}}}

\newcommand{\ket}[1]{\ensuremath{\big| #1 \big\rangle}}

\newcommand{\e}[0]{\ensuremath{\epsilon}}

\newcommand{\q}[1]{\ensuremath{\langle #1 \rangle}}

\newcommand{\const}{\mathrm{const}}

\usepackage{xfrac}

\newcommand{\pa}[1]{\left(#1\right)} 
\newcommand{\bra}[1]{\left[#1\right]}

\newcommand{\mat}[1]{\begin{matrix}#1\end{matrix}} 
\newcommand{\pmat}[1]{\pa{\mat{#1}}}

\newcommand{\pfrac}[2]{\pa{\frac{#1}{#2}}}

\renewcommand{\Re}{\mathrm{Re}\,}

\begin{document}

\title{Degenerate Bose gases with uniform loss
}
\author{Pjotrs Gri\v{s}ins}
\email{pjotrs.grisins@unige.ch}
\affiliation{Department~of~Quantum~Matter~Physics, University~of~Geneva, 24~Quai~Ernest~Ansermet, 1211~Geneva, Switzerland}
\affiliation{Vienna~Center~for~Quantum~Science~and~Technology, Atominstitut, Stadionallee~2, 1020~Vienna, Austria}

\author{Bernhard Rauer}
\affiliation{Vienna~Center~for~Quantum~Science~and~Technology, Atominstitut, Stadionallee~2, 1020~Vienna, Austria}

\author{Tim Langen}
\altaffiliation[Present address: ]{JILA, NIST~\&~Department~of~Physics, University~of~Colorado, 440~University~Ave, Boulder, Colorado~80309.}
\affiliation{Vienna~Center~for~Quantum~Science~and~Technology, Atominstitut, Stadionallee~2, 1020~Vienna, Austria}

\author{J\"org Schmiedmayer}
\affiliation{Vienna~Center~for~Quantum~Science~and~Technology, Atominstitut, Stadionallee~2, 1020~Vienna, Austria}

\author{Igor E. Mazets}
\affiliation{Vienna~Center~for~Quantum~Science~and~Technology, Atominstitut, Stadionallee~2, 1020~Vienna, Austria}
\affiliation{Wolfgang~Pauli~Institute, Oskar-Morgenstern-Platz~1, 1090~Vienna, Austria}

    \begin{abstract} \noindent
    We theoretically investigate a weakly-interacting degenerate Bose gas coupled to an empty Markovian bath. We show that in the universal phononic limit the system evolves towards an asymptotic state where
an emergent temperature is set by the quantum noise of the outcoupling process. For situations typically encountered in experiments, this mechanism leads to significant cooling. Such dissipative cooling supplements conventional evaporative cooling and dominates in settings where thermalization is highly suppressed, such as in a one-dimensional quasicondensate.
		\end{abstract} 
		
\pacs{03.75.Kk, 05.70.Ln, 03.65.Yz}

    \maketitle

\section{Introduction}

Engineering dissipation and driving protocols in interacting quantum many-body systems is an important emerging area of out-of-equilibrium physics. On the theoretical side, it has unveiled a series of novel quantum phenomena,  from topological states of fermions \cite{Budich2014} and the establishment of long-range order of a Bose-Einstein condensate (BEC) in an optical lattice \cite{Diehl2008}, to the  dissipative preparation of entangled states \cite{Weimer2010} and dissipative quantum computations \cite{Verstraete2009}. On the experimental side, dissipation, for instance, has been used to create strongly correlated states of matter \cite{Syassen2008} and to study the dynamics of open quantum systems \cite{Barontini2013}.

In the present article we develop a general model for dissipative ultracold bosonic gas, where the dissipation is based on spatially uniform and
coherent atomic loss from a BEC into a continuum of
free single-particle modes.
In contrast to existing studies of atom lasers \cite{Japha2002,Japha1999}, in the present work we concentrate not on the coherence properties of the outcoupled atoms, but on the dissipation-driven evolution of the remaining ones.
Our model is also different from conventional driven-dissipative models discussed in the literature \cite{Sieberer2015,Maghrebi2015a} as in our case there is no driving; consequently, there is no steady state: the system asymptotically approaches the true vacuum. We are interested in the out-of-equilibrium transient dynamics during the evolution to this trivial final state.

\begin{figure}
	\centering
	\includegraphics[width=0.5\linewidth]{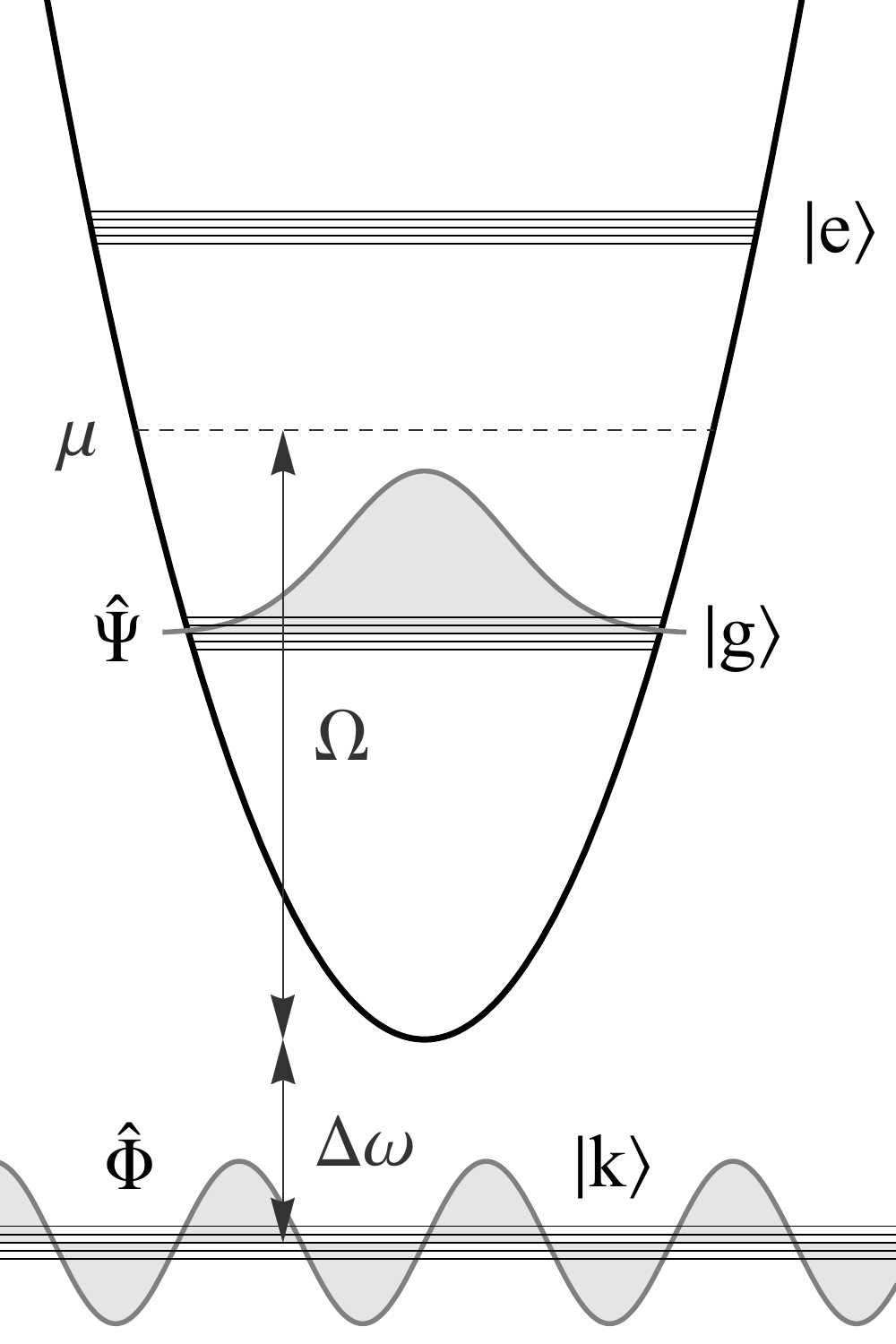}
	\caption{Schematic of the considered setup. Interacting degenerate bosonic atoms in the transversal ground state of a harmonic trap $\ket{g}$ are outcoupled to a continuum of free modes $\ket{k}$ using a microwave or an rf-field with a Rabi frequency $\Omega$ and detuning $\Delta \omega$. The interactions between the atoms in $\ket{g}$  are manifested in the mean-field shift $\mu \equiv g \q{\hat\Psi^\dag \hat \Psi}$, where $g=\const$ is the self-interaction strength. The mean-field shift is still considerably smaller than the energy of the first transversally excited level $\ket{e}$. The kinetic energy in the longitudinal direction (inside the plane of the paper), represented by the fine structure of the bands, is much smaller 
		than any other relevant energy scale.
	}
	\label{fig:1}
\end{figure}

A specific realization of dissipation that we will concentrate on is a one-dimensional (1D) degenerate Bose gas, where trapped atoms are coupled to an untrapped state with a radio-frequency or microwave transition, Figure~\ref{fig:1}.  This is closely related to a recent experimental study of cooling in a 1D quantum gas \cite{Rauer2015}.
Although reminiscent
of standard evaporative cooling \cite{Ketterle1996,Surkov1996,Luiten1996}, the process observed there
is distinctly different in that it neither relies on energy-selective outcoupling nor conventional re-thermalization.

The paper is organized as follows: 
 in Section \ref{sec:mark} we introduce the model and derive an effective stochastic Gross-Pitaevskii equation in the Markovian approximation. In Section \ref{sec:lin} we linearize it and focus on the experimentally relevant quasi-stationary dissipation process in the low-energy phononic limit. In Section \ref{sec:5} we show that in this regime the elementary excitations are in a thermal state with a time-dependent effective temperature. Finally, we obtain experimentally relevant scaling
laws for this temperature and find an asymptotic dissipative state, which emerges at long timescales.

%
%
%
%

\section{Markovian dissipation}\label{sec:mark}

 
 We start with a degenerate Bose gas in a trap, where the radial confinement is orders of magnitute larger than the longituninal one. For instance, this situation can be realized on atom chips \cite{Kruger2010,Manz2010,Reichel2011}. In this case the single-particle wavefunction factorizes into radial and longitudinal components as $\psi(x,y,z) = \psi_\perp(x,y)\,\psi_\parallel(z)$. Due to the strong transversal confinement, the gas is in the radial ground state, which is represented by a Gaussian wavefunction
\begin{equation}
\label{eq:Gaussian}
 \psi _\perp (x,y) =\dfrac{1}{\sqrt{\pi} \sigma} e^{-\frac{x^2+y^2}{2\sigma ^2}},
\end{equation}
where  $\sigma =1/\sqrt{m\omega _\perp }$ is the width of the ground state, $m$ is the mass of bosonic particles, $\omega _\perp $ is the fundamental frequency of the radial trapping potential, and we set $\hbar= k_B = 1$.
The linear density of atoms $n$
is assumed to be small enough ($n a_s \ll 1,$ where $a_s$ is the scattering length), so that we can neglect the mean-field broadening~\cite{Salasnich2002}.

To be able to derive general analytical results, we neglect the interactions of trapped atoms with the atoms in the 
untrapped state, therefore assuming 
that the eigenfunctions of the latter are plane waves in the $x$ and $y$ dimensions.
 This is justified in 1D by the fact that outcoupled particles leave the condensate region at a short time scale $\sim 1/\omega_\perp$. Mean-field repulsion from the condensate creates a potential peak at the center of the trap, which additionally accelerates the outcoupled particles. However this effect does not change the physics qualitatively \cite{Japha2002}.
 We also neglect gravity, which would render the transversal eigenfunctions in the direction of free fall to be Airy functions. Both before-mentioned effects can be easily taken into account numerically when analyzing particular experimental implementations.

The 1D field operators for trapped and  untrapped atoms having the momentum $(k_x,k_y)$ in the radial directions are, respectively, $\hat \Psi = \hat \Psi(z,t)$ 
and $\hat \Phi _{k_x,k_y} = \hat \Phi _{k_x,k_y}(z,t)$. The coupled set of equations for them reads 
\begin{widetext}
\begin{subequations}	
\begin{align} 
i \frac \partial {\partial t}\hat \Psi &= -\frac 1{2m} \frac {\partial ^2}{\partial z^2} \hat \Psi +g \hat \Psi ^\dag \hat \Psi 
\hat \Psi 
+ \sum  _{k_x,k_y}\kappa ^* _{k_x,k_y}\hat \Phi _{k_x,k_y}, 
\label{UT.2.1} \\ 
i \frac \partial {\partial t}\hat \Phi _{k_x,k_y}&=\left( \frac {k_x^2+k_y^2}{2m} -\Delta \omega \right) \hat \Phi _{k_x,k_y} 
-
\frac 1{2m} \frac {\partial ^2}{\partial z^2}\hat \Phi _{k_x,k_y}+\kappa  _{k_x,k_y}\hat \Psi, 
\label{UT.2.2} 
\end{align} 
\end{subequations}
\end{widetext}
where $g$ is the self-interaction strength.

Here the coupling between the trapped and untrapped fields is given by the overlap of the respective single-particle wavefunctions
\begin{equation} 
\kappa  _{k_x,k_y} =\frac \Omega {\sqrt{A}} \int dx \int dy\, e^{-ik_xx-ik_yy} \psi _\perp (x,y) , 
\label{UT.3}
\end{equation} 
where $A$ is the quantization area in the $(x,y)$-plane, and 
$\Omega $ is the Rabi frequency of the microwave- or rf-driven transition. The detuning is denoted by $\Delta \omega $ (see Figure~\ref{fig:1} for the energy level diagram). 

Since \eqref{UT.2.2} is linear, we may express $\hat \Phi _{k_x,k_y}(t)$ through the formal solution 
\begin{align}
	\hat \Phi _{k_x,k_y}(t) =& e^{it\bra{ \Delta \omega + \frac 1{2m}\pa{ \frac {\partial ^2}{\partial z^2} - k_x^2-k_y^2 }}}\hat \Phi _{k_x,k_y}(0)-
	\nonumber\\
	&-i \kappa _{k_x,k_y}\int_{0}^{t}dt^\prime \, e^{i(t-t^\prime )\bra{ \Delta \omega + \frac 1{2m}\pa{ \frac {\partial ^2}{\partial z^2} - k_x^2-k_y^2}  }}\hat \Psi (t^\prime )
	,   \label{form.sol}
\end{align} 
where $\hat \Phi _{k_x,k_y}(0)$ are the initial conditions for $\hat \Phi _{k_x,k_y}$ at $t=0$. Substituting (\ref{form.sol}) into \eqref{UT.2.1}, we 
obtain
\begin{align} 
i \frac \partial {\partial t}\hat \Psi = &-\frac 1{2m} \frac {\partial ^2}{\partial z^2} \hat \Psi +g \hat \Psi ^\dag \hat \Psi 
\hat \Psi 
 -\nonumber\\
 &-i \int _0^tdt^\prime \, F(t-t^\prime )\,
e^{i(t-t^\prime )\pa{\Delta \omega +\frac{1}{2m}\frac{\partial ^2}{\partial z^2}}}\hat \Psi (t^\prime ) +\hat \varsigma (t), 
\label{UT.4} 
\end{align} 
where the quantum noise term is given by
\begin{equation} 
\hat \varsigma (t)=\sum  _{k_x,k_y}\kappa ^* _{k_x,k_y}e^{it\bra{ \Delta \omega + \frac{1}{2m} \pa{\frac{\partial^2}{\partial z^2} -k_x^2-k_y^2}} }
\hat \Phi _{k_x,k_y}(0), 
\label{UT.5} 
\end{equation} 
and the kernel of the integral term is 
\begin{align} 
F(\tau ) &= \sum  _{k_x,k_y}|\kappa  _{k_x,k_y}|^2 e^{-\frac{i(k_x^2+k_y^2)\tau}{2m}} \equiv \nonumber\\ &\equiv
A \int \frac {dk_x}{2\pi } \int \frac {dk_y}{2\pi } \, |\kappa  _{k_x,k_y}|^2 e^{-\frac{i(k_x^2+k_y^2)\tau}{2m}} \,  . ~
\label{UT.6}
\end{align} 
We explicitly indicate the time argument of the fields $\hat \Psi $ and $\hat \Phi _{k_y,k_z} $ in Eqs. (\ref{UT.4}) and 
(\ref{UT.5}), respectively, when it differs from $t$. 
Assuming that the atomic interactions do not affect strongly the transverse profile of the trapped atomic cloud and that the latter remains Gaussian, from \eqref{eq:Gaussian} and \eqref{UT.3} we obtain
\begin{equation} 
|\kappa _{k_x,k_y}|^2 =\frac {2\pi \Omega ^2}{Am\omega _\perp} \, e^{-\frac{k_x^2+k_y^2}{m\omega _\perp}},\label{UT.8new}
\end{equation}
and substituting \eqref{UT.8new} into \eqref{UT.6} arrive at
\begin{equation} 
F(\tau ) =\frac {\Omega^2}{1+i\omega _\perp \tau /2} . 
\label{UT.7}
\end{equation} 

In general the dissipation and noise terms are explicitly non-Markovian and non-local. However we can simplify them by considering times $t \gg 1/\omega _\perp$, the  characteristic timescale set by  (\ref{UT.7}). We assume that all the relevant energy scales of the system, given by  the temperature, the chemical potential and the Rabi frequency $\Omega$ are well below $\omega _\perp$. This allows us to pull $\hat \Psi (t^\prime )$ out of the integral in the r.h.s. of  
(\ref{UT.4}), replacing $t^\prime $ by $t$, and to neglect the small kinetic energy in the longitudinal direction
\begin{equation} 
e^{\frac {i(t-t^\prime )}{2m}\frac {\partial ^2}{\partial x^2}} \approx 1,  
\label{UT.8} 
\end{equation} 
which is an analogue of the Thomas-Fermi approximation.
The long-time condition $t\gg 1/\omega _\perp $ allows us to substitute the upper limit of the time integral in 
the r.h.s. of  (\ref{UT.4}) by $\infty $. The imaginary part of $\int _0 ^t d\tau\, F(\tau ) \exp (i\Delta \omega \tau )$ 
renormalizes the energy of the trapped state and can be incorporated into $\Delta \omega $. The real part determines the 
loss rate 
\begin{align} 
\gamma &= \Omega ^2\, \mathrm{Re} \int _0^t d\tau\, \frac {\exp (i\Delta \omega \tau )}{1+i\omega _\perp \tau /2} 
 \approx \nonumber\\&\approx
\frac {2\pi \Omega ^2}{\omega _\perp } \exp \left( -\frac {2\Delta \omega }{\omega _\perp } \right) \Theta (\Delta \omega ), 
\label{UT.9} 
\end{align}  
where $\Theta (\Delta \omega )$ is Heaviside's step function, and in the last line we took $t \gg 1/\Delta \omega$. Hence the non-Markovianity time of our system is given by $\tau_M \sim \max(1/\omega_\perp, 1/\Delta \omega)$.

As a result we arrived at a local and Markovian constant dissipation term at $t>\tau_M$, assuming that the detuning $\Delta \omega$ is positive and is held constant with respect to the decaying chemical potential of the remaining atoms.

What conserns the dissipation-induced noise, we consider an empty bosonic bath, so we have only vacuum fluctiuations of the untrapped-atom field: 
\begin{align} 
\langle \hat \Phi ^\dag _{k_x^\prime ,k_z^\prime }(z^\prime ,0) \,\hat \Phi _{k_x ,k_y }(z ,0)\rangle & =0, 
\nonumber\\
\langle \hat \Phi _{k_x ,k_y }(z ,0)\,\hat \Phi ^\dag _{k_x^\prime ,k_y^\prime }(z^\prime ,0) \rangle & = 
\delta _{k_x\, k_x^\prime } \delta _{k_y\, k_y^\prime }\delta (z -z^\prime ). 
\label{UT.12} 
\end{align} 
Then from \eqref{UT.5}, \eqref{UT.9} and \eqref{UT.12}  we obtain the correlators for the quantum noise 
\begin{align} 
\langle \hat \varsigma ^\dag (z^\prime ,t^\prime )\,\hat \varsigma (z,t)\rangle &=0, 
\\
\langle \hat \varsigma (z,t)\,\hat \varsigma ^\dag (z^\prime ,t^\prime )\rangle &=  \delta (z-z^\prime )  
\frac {\Omega ^2\exp [i\Delta \omega (t-t^\prime ) ]}{1+i \omega _\perp (t-t^\prime )/2}. 
\label{UT.14} 
\end{align}
The fluctuation-dissipation theorem manifests itself in the relation between the noise corellator and the dissipation rate $\gamma$ according to
\begin{equation} 
\int _{-\infty }^\infty dt^\prime \, \langle \hat \varsigma (z,t)\hat \varsigma ^\dag (z^\prime ,t^\prime )\rangle =
2\gamma \delta (z-z^\prime ) . 
\label{UT.15} 
\end{equation} 

Again assuming that the non-Markovianity time $\tau_M$ is smaller than any relevant time scale of the system, we can approximate the Lorentzian of \eqref{UT.14} with a delta-function
\begin{equation}
\langle \hat \varsigma(z,t) \hat\varsigma^\dag(z',t')\rangle = 2\gamma\,\delta(z-z')\,\delta(t-t').
\label{noise2}
\end{equation}

Physically the assumed Markovianity is due to the fact that after a microwave- or rf-induced transfer to the untrapped state the atom quickly leaves the trap (cf. free evolution of a Gaussian \eqref{eq:Gaussian}), so after the time $t \sim
1/\omega_\perp$ it is on average too far from the trap  and has basically no probability to absorb another photon and return to the condensate~\footnote{In the opposite limit of long non-Markovianity time the atoms would perform Rabi-type oscillations between the condensate and the untrapped state, but this regime is out of scope of the current article.}.

As a result we obtain the Markovian and local dissipative equation of motion for the trapped-atom field operator 
\begin{equation} 
i \frac \partial {\partial t}\hat \Psi = -\frac 1{2m} \frac {\partial ^2}{\partial z^2} \hat \Psi +g \hat \Psi ^\dag \hat \Psi 
\hat \Psi 
 -i\gamma  \hat \Psi +\hat \varsigma,
\label{UT.10} 
\end{equation}
which is the starting point for all derivations of the next sections. 

For reference, in the mean-field aproximation \eqref{UT.10} becomes the standard Gross-Pitaevskii equation with an additional dissipative term
\begin{equation} 
	i \frac \partial {\partial t} \Psi_\mathrm{mf} = -\frac 1{2m} \frac {\partial ^2}{\partial z^2} \Psi_\mathrm{mf} +g |\Psi_\mathrm{mf}|^2 
	 \Psi_\mathrm{mf} 
	  -i\gamma  \Psi_\mathrm{mf},
	\label{UT.10mf} 
\end{equation}
where the mean-field treatment amounts to solving \eqref{UT.10} for expectation values of field operators, $\hat \Psi(z,t) \rightarrow \q{\hat \Psi(x,t)} = \Psi_\mathrm{mf}(z,t)$, and neglecting the quantum noise term $\hat \varsigma \rightarrow \q{\hat \varsigma} = 0$.

\section{Linearized analysis}\label{sec:lin}
\subsection{Bogoliubov theory}

We start with the phase--density representation \cite{Mora2003, Petrov2000} of the field operator
\begin{align}
\hat \Psi(z,t)& = e^{i \hat \theta(z,t)} \sqrt{n(t) + \delta \hat n(z,t)},\label{phase-density}
\end{align}
where $\hat \theta = \hat \theta(z,t)$ is the phase operator, $ \delta \hat n = \delta \hat n(z,t)$ is the density fluctuation operator, and $n = n(t) = n_0\, e^{-2\gamma t}$ is an exponentially decaying mean density.

Substituting the field operator \eqref{phase-density} into the equation of motion \eqref{UT.10}, and linearizing the latter with respect to the small density fluctuations and phase gradients, we acquire the equations of motion for the phase and density operators
\begin{align} 
\frac{\partial}{\partial t} \hat \theta &= -\left( g -\frac 1{4mn} \frac {\partial ^2}{\partial z^2} \right) \delta \hat n +
\frac {\hat s +\hat s^\dag  }{2\sqrt{n}} ,     \nonumber \\ 
\frac{\partial}{\partial t} \delta \hat n &=-\frac {n}m \frac {\partial ^2}{\partial z^2} \hat \theta -2\gamma\, \delta \hat n+i
\sqrt{n}(\hat s -\hat s ^\dag),     \label{UT.18} 
\end{align}
where 
\begin{equation}
\hat s = \hat s(z,t) = \hat \varsigma(z,t)\,e^{-i \hat \theta(z,t)},
\label{noise1}
\end{equation}
and we take into account that $\hat \varsigma$ and $\hat \theta$ commute.

We note that the phase-density representation is valid for  1D, 2D and 3D degenerate bosonic gases in both regimes of a true BEC and a quasi-BEC. In the case of a true BEC the phase fluctuations are also suppressed, so the former equations can be further simplified by expanding the phase exponential $e^{i\hat \theta} \approx 1 + i\hat \theta$. However we do not make this approximation to keep the discussion applicable to lower dimensions, where the phase fluctuations may be strong.

To find the elementary excitations of the system we first perform an instantaneous unitary transformation to an emergent bosonic basis \cite{Mora2003}
\begin{equation}
\mat{
	\hat \varphi = \dfrac{\delta \hat n}{2\sqrt{n}} + i \sqrt{n}\,\hat  \theta, \qquad &
	\delta \hat n = \sqrt{n} (\hat \varphi + \hat \varphi^\dag), \\
	\hat  \varphi^\dag = \dfrac{\delta \hat n}{2\sqrt{n}} - i \sqrt{n} \, \hat \theta, \qquad &
	\hat \theta = \dfrac{1}{2i\sqrt{n}} (\hat \varphi- \hat \varphi^\dag),
}
\end{equation}
which leads to
\begin{gather}
i\, \partial_t \hat \varphi + \frac 1{2m} \partial_{zz} \hat \varphi - g n (\hat \varphi + \hat \varphi^\dag) + i \gamma\, \hat \varphi + \hat s= 0.\label{eq21}
\end{gather}

In the following we concentrate on the 1D case in a box of length $L$ with periodic boundary conditions. Later we will use the local density approximation to infer the properties of a trapped gas from the untrapped one.

 After the Fourier transformation
\begin{align}
f(z) &= \dfrac{1}{\sqrt{L}} \sum_k f_k e^{ikz},
\nonumber\\ f(z)^\dag &= \dfrac{1}{\sqrt{L}} \sum_k f_{k}^\dag e^{-ikz} = \dfrac{1}{\sqrt{L}} \sum_k f_{-k}^\dag e^{ikz},
\end{align}
the equation \eqref{eq21} reads
\begin{gather}
i\, \partial_t \hat \varphi_k - \dfrac{k^2 \hat \varphi_k}{2m} - g n (\hat \varphi_k + \hat \varphi_{-k}^\dag) + i \gamma\, \hat \varphi_k + \hat s_k= 0,
\nonumber\\
i\, \partial_t \hat \varphi^\dag_{-k} + \dfrac{k^2 \hat \varphi_{-k}^\dag}{2m} + g n (\hat \varphi_{k} + \hat \varphi^\dag_{-k}) + i \gamma\, \hat \varphi_{-k} - \hat s^\dag_{-k}= 0,
\end{gather}
or in the vector form
\begin{gather}
i\partial_t  \Phi - H \Phi + i \gamma  \Phi + S = 0,
\end{gather}
where 
\begin{gather}
\Phi = \pmat{\hat \varphi_k \\ \hat \varphi_{-k}^\dag},
\quad
S = \pmat{\hat s_{k} \\ - \hat s^\dag_{-k}},
\nonumber\\
H = \pmat{\sfrac{k^2}{2m}+gn&gn\\-gn&-\sfrac{k^2}{2m}-gn}.
\end{gather}

The Hamiltonian $H$ can be diagonalized using the standard Bogoliubov rotation (we set $u,v = u_k,v_k$ to be real for convenience), given by
\begin{gather}
\Phi =  P X,
\nonumber\\
X = \pmat{\hat\chi_k \\  \hat\chi^\dag_{-k}}, \quad
P = \pmat{u&-v\\-v&u}, \quad
P^{-1} = \pmat{u&v\\v&u},
\nonumber\\
D =  P^{-1}  H  P = \pmat{\e_k&0\\0&-\e_k},\quad
\e_k = \sqrt{E_k\pa{E_k + 2\mu}},
\nonumber\\
\mu = g n, \quad E_k = \frac{k^2}{2m},
\nonumber\\
u^2 - v^2 = 1, \quad
u^2, v^2 = \dfrac{1}{2\e_k}\pa{\pm \e_k + E_k + \mu},
\nonumber\\
u \pm v = \pa{\dfrac{E_k}{\e_k}}^{\mp \sfrac 12}, \quad
2 u v = \frac{\mu}{\e_k},
\quad
u^2 + v^2 = \dfrac{E_k + \mu}{\e_k},
\end{gather}
where $\hat \chi_k$ are Bogoliubov modes, and $\mu$ is the time-dependent mean-field shift (chemical potential).
This leads to
\begin{gather}
i (P^{-1} \partial_tP)X + i \partial_t X - D X + i \gamma X + P^{-1} S = 0,
\end{gather}
and to the equation of motion for the components
\begin{gather}
i(v \partial_t u - u \partial_t v) \hat \chi_{-k}^\dag 
+i \partial_t \hat \chi_k - \e_k \hat \chi_k +
\nonumber\\
+ i\gamma \hat \chi_k + u_k \hat s_k - v_k \hat s^\dagger_{-k} = 0,
\label{Bogoliubov}
\end{gather}
where we used $u \partial_t u - v \partial_t v = 0$.

Analyzing \eqref{Bogoliubov} we see that it differs from the standard equation of motion for Bogoliubov quasiparticles $i\partial_t \hat \chi_k = \epsilon_k \hat \chi_k$ in three important aspects, as it includes

1) a non-adiabatic contribution $\pa{v \partial_t u - u \partial_t v}$ due to the decreasing mean density $n$ and corresponding change in the mode energy $\e_k$ and Bogoliubov coefficients $u_k, v_k$;

2) an adiabatic loss term $i\gamma \hat \chi_k$, leading to exponential decay of the number of elementary excitations in each momentum mode;

3) and a squeezed quantum noise term $(u_k \hat s_k - v_k \hat s^\dagger_{-k})$, where the squeezing is due to the transformation from the real particle basis into the Bogoliubov basis.

Eq. \eqref{Bogoliubov} cannot be analytically solved in full generality, however it can be conveniently analyzed in the experimentally relevant limits.

\subsection{Non-adiabatic corrections}

Let us first consider non-adiabatic corrections.
After some algebra we get $v \partial_t u - u \partial_t v = \frac{\gamma}{2+E_k/\mu(t)} = \frac{\tilde\gamma_k(t)}{2}$, which is a monotonously decaying function of time (recall that $\mu(t)$ decays exponentially), bounded from above by $\tilde\gamma_k(t)\leqslant\gamma$, which becomes equality in the phononic limit $E_k \ll \mu(t)$.

Considering for a moment the mean-field theory and disregarding the quantum noise terms in \eqref{Bogoliubov}, we see that the non-adiabatic terms mix $\pm k$ field components.
\begin{align}
i \partial_t \pmat{\hat\chi_k \\  \hat\chi^\dag_{-k}} 
=
\pmat{ \e_k(t) - i\gamma & -i \tilde \gamma_k(t)/2 \\ -i \tilde \gamma_k(t)/2 & -\e_k(t) - i\gamma} \pmat{\hat\chi_k \\  \hat\chi^\dag_{-k}} + \ldots,
\label{eq:non-adiabatic}
	\end{align}
where $(\ldots)$ represents the omitted noise terms. Assuming that the dissipation is slow enough that $|\e_k(t)|$ is almost constant during one period of oscillation $\tau_k = 2\pi/\e_k$, we can diagonalize again this `dissipation-dressed Hamiltonian' by applying a quasistationary approximation. The instantaneous spectrum acquires a diffusive part, the complex energies being \begin{equation}
\tilde \e_k(t) = \pm \sqrt{\e_k(t)^2 - \frac 14 \tilde \gamma^2_k(t)} - i\gamma.
\label{eq:dispersion1}
\end{equation}

\begin{figure}
	\centering
	\includegraphics[width=0.9\linewidth]{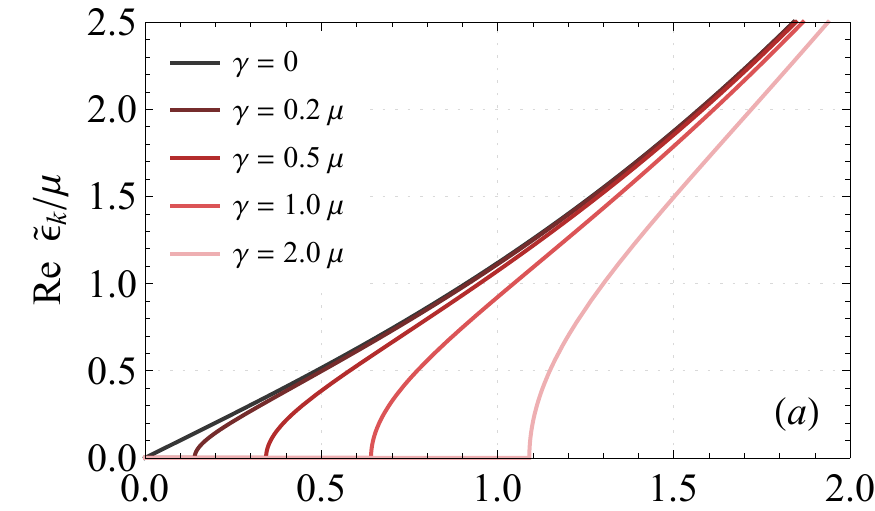}
	\includegraphics[width=0.9\linewidth]{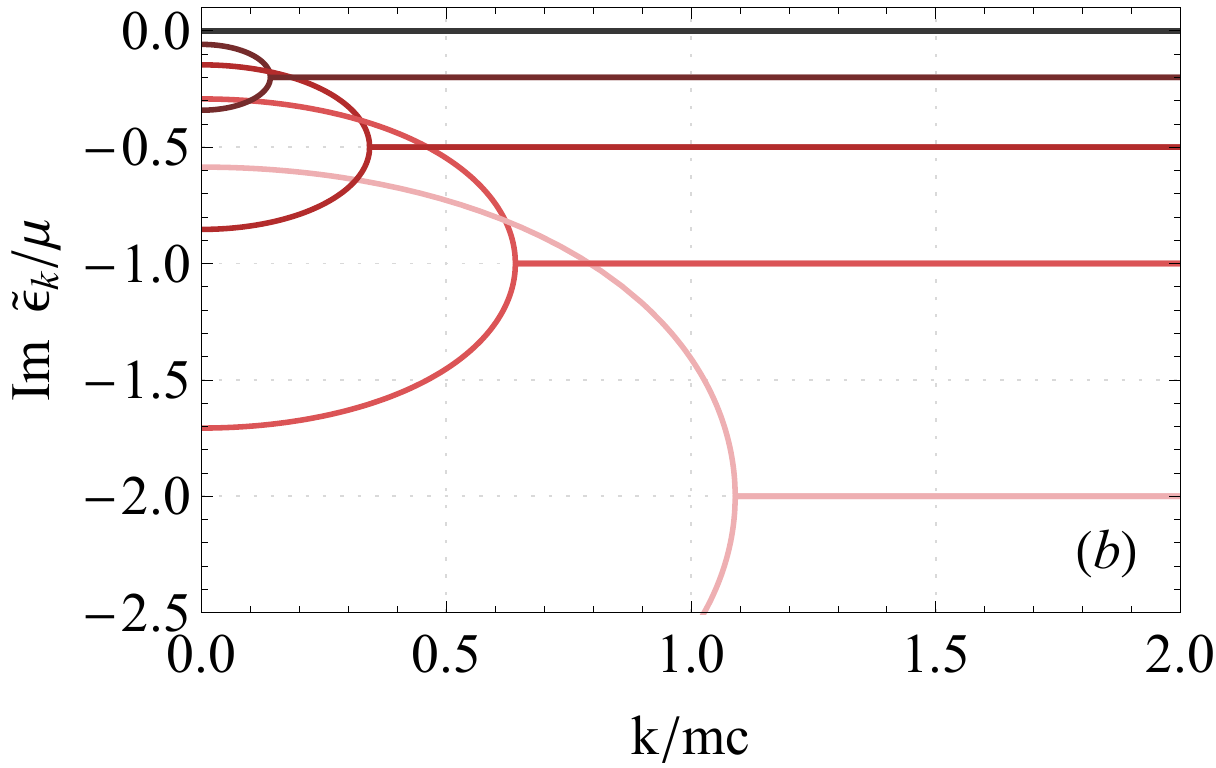}
	\caption{Approximate dispersion curves for a one-dimensional dissipative quasicondensate. The panels show real (a) and imaginary (b) parts of the quasiparticle energy $\tilde \e_k$ (in units of the chemical potenial $\mu$) as a function of momentum (in units of the inverse healing length $1/\xi = mc$) and the dissipation strength. 
		The quasiparticle decay rate is the negative imaginary part of the energy.
		Diffusive modes appear when the real part of the dispersion touches zero. The two diffusive modes, density- and phase-like, have different decay rates, which is represented by two branches in panel (b). Note that $\gamma$ is the dissipation rate of the order parameter, and not of the density, namely $n(t) = n_0\, e^{-2\gamma t}$.
	}
	\label{fig:2}
\end{figure}

The solutions to \eqref{eq:dispersion1} are plotted in Figure~\ref{fig:2} for $t=0$. At low dissipation rates $\gamma \ll \mu$ we recover the standard Bogoliubov dispersion relation with a small imaginary component, which corresponds to a finite lifetime of the quasiparticles. In this limit the quasiparticle mode occupation numbers decay at the same rate as the mean density.

In the opposite limit of strong dissipation the modes become non-propagating ($\Re \tilde \epsilon_k = 0$), but diffusive.
When the probed length scales are dominated by diffusive modes, any local perturbation will not lead to a light-cone-like spread of correlations \cite{Langen2013}, but will smoothly decay similar to the solutions of the heat equation.

However we note that the diffusive modes may not be  easily accessible in a purely dissipative system, as at large~$\gamma$ the mean density may decay too quickly for any observable effects. However these modes may become observable if one designs a pumping scheme to counterbalance the density loss, in a similar spirit as it is done with exciton-polariton condensates \cite{Wouters2009,Carusotto2013}.

In any case, in current 1D quasicondensate experiments \cite{Rauer2015} the dissipation rate is $\gamma \sim 10^{-3} \mu$, and for  experimentally accessible momenta $k \gtrsim 1/L \gtrsim 0.1\, mc$ (here $L$ is the length of the cloud and $c = \sqrt{\mu/m}$ is the speed of sound)
the contribution of the non-adiabatic term is negligibly small. So in the following we consider dissipation to be adiabatic.

\subsection{Quantum noise}\label{sec:noise}

In the adiabatic limit, where $\gamma$ is much less than any other relevant energy scale, Eq.~\eqref{Bogoliubov} reads
\begin{gather}
i \partial_t \hat \chi_k - \e_k \hat \chi_k 
+ i\gamma \hat \chi_k + u_k \hat s_k - v_k \hat s^\dagger_{-k} = 0,
\label{2}
\end{gather}
and has the solution
\begin{gather}
\hat \chi_k(t) = e^{-(i \e_k  + \gamma) t}\, \hat \chi_k(0) +
\nonumber\\
+ i \int_0^t e^{-(i\e_k + \gamma)(t-t')} [ u_k \hat s_k(t') - v_k \hat s^\dagger_{-k}(t')]\,dt'.
\end{gather}
Equal-time normal $n_k(t) = \q{\hat \chi_k(t)^\dag \hat \chi_k(t)}$ and anomalous $m_k(t) = \q{\hat \chi_k(t) \, \hat \chi_{-k}(t)}$ correlators evolve according to
\begin{gather}
n_k(t) = e^{-2\gamma t} n_k(0) + 
\nonumber\\
+
\iint^t_0 v_k^2\, \q{\hat s_{-k}(t'') \hat s^\dagger_{-k}(t')}\, e^{-2 \gamma t + i \e_k(t'-t'') + \gamma(t'+t'')} \, dt'\ dt'',
\nonumber\\
m_k(t) = e^{-2(i\e_k+\gamma) t} m_k(0) + 
\nonumber\\
+\iint^t_0 u_k v_k\, \q{\hat s_{k}(t'') \hat s^\dagger_{k}(t')}\, e^{-2(i\e_k+ \gamma) t + (i \e_k + \gamma)(t'+t'')} \, dt'\ dt'',
\label{nm1}
\end{gather}
where we have taken into account that 
\begin{align}
\q{\hat s^\dagger_{k}(t'') \hat s_{k'}(t')} &= 0,
\\
\q{\hat s_{k}(t'') \hat s^\dagger_{k'}(t')} &= 0 \quad \mathrm{for} \quad k\neq k'. 
\end{align}
Noticing the statistical independence of $\hat \varsigma$ and $\hat \theta$, and using \eqref{noise1} and \eqref{noise2} we get the remaining quantum noise correlator
\begin{gather}
\q{\hat s_{k}(t) \hat s^\dagger_{k}(t')} =
\nonumber\\
= \frac 1L
\iint dz\, dz'\, \q{\hat \varsigma(z,t) \hat \varsigma(z',t')} \q{e^{-i\hat \theta(z,t) +i\hat \theta(z',t')}} e^{-ik(z-z')}=
\nonumber\\
= 2\gamma\,\delta(t-t').
\label{noise3}
\end{gather}

In this derivation the Markovianity and locality of the noise was essential to ensure that $\q{e^{-i\hat \theta(z,t) +i\hat \theta(z',t')}} = 1$. If there exists residual non-Markovianity, the phase fluctuations in space and time, which are especially strong in 1D, will reduce this correlator. So we may expect that the quantum noise influence is reduced in a general non-Markovian case. 

Substituting \eqref{noise3} into \eqref{nm1} we get
\begin{align}
n_k(t) =& e^{-2\gamma t} n_k(0) +
2\gamma\int^t_0 v^2_k(t')\,e^{-2\gamma(t-t')}\, dt',
\nonumber\\
m_k(t) =& e^{-2(i\e_k+\gamma) t} m_k(0) +
\nonumber\\ 
&+2\gamma \int^t_0 u_k(t') v_k(t')\,e^{-2(i\e_k + \gamma)(t-t')}\, dt'.
\label{nm2}
\end{align}

\begin{figure}
	\centering
	\includegraphics[width=0.9\linewidth]{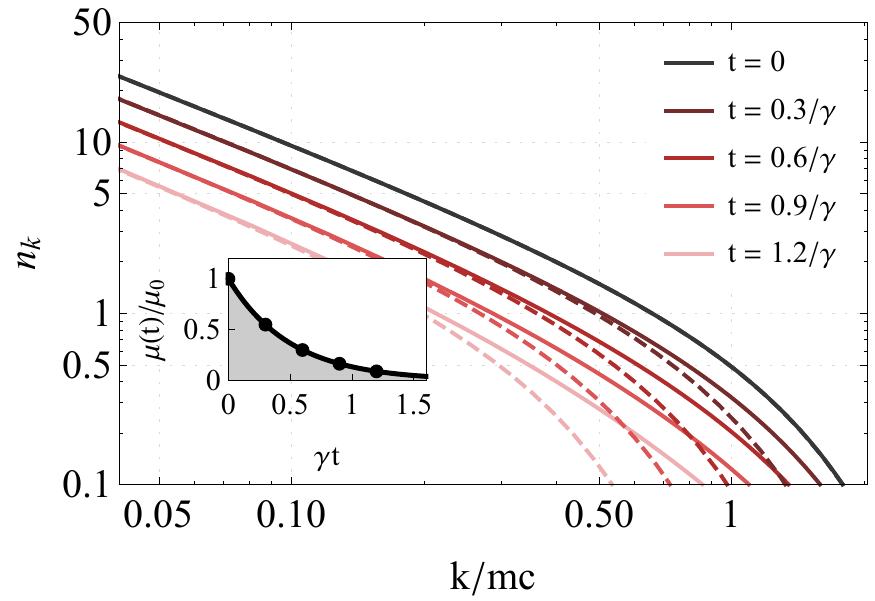}
	\caption{Log-log plot of the evolution of the Bogoliubov modes' ocupation numbers $n_k$ as a function of momentum $k$ (in units of the inverse healing length $1/\xi = mc$) and time $t$ (solid lines, in units of the inverse dissipation rate $1/\gamma$) for initial thermal distribution at temperature $T_0 = \mu_0$. It is easy to see the two limits of the full Bose-Einstein distribution function: the phononic Rayleigh-Jeans limit ($k\lesssim mc$) and the particle-like Boltzmann limit ($k\gtrsim mc$).
		 The dashed lines represent thermal distributions at temperatures, top to bottom, $T_\eff(t)/\mu_0 \approx \{1.00,0.55,0.30,0.17,0.09\}$, fitted as to
		  agree with the calculated values in the phononic regime. Note that although the time-evolved distributions are clearly non-thermal in their high-energy tails, they however agree well with thermal predictions in the low-energy part of the spectrum, allowing to introduce an effective temperature for phononic modes.  
Inset: time evolution of the chemical potential $\mu(t) = \mu_0\, e^{-2\gamma t}$,
		   solid line, in comparison with the fitted effective temperatures $T_\eff(t)$, dots. As explained in Section~\ref{sec:5}, in this special case $T(t)=\mu(t)$.}
	\label{fig:3}
\end{figure}

Numerical solutions to \eqref{nm2} for experimental parameters of \cite{Rauer2015} are presented in Figure~\ref{fig:3}, with the assumptions of independence of Bogoliubov modes, and setting the initial state to be a true thermal equilibrium at a temperature $T_0 = \mu_0$. The initial condition for the correlators are
\begin{align}
n_k(0) &= (e^{\e_k/T_0}-1)^{-1},
	\nonumber\\
	m_k(0) &= 0.
	\label{initial}
\end{align}

In Figure~\ref{fig:3} we see that although the quasiparticle occupation numbers deviate strongly from the predictions of the thermal Bose-Einstein distribution for high-momenta particle-like states, the low-energy phononic excitations  agree very well with the Rayleigh-Jeans classical equipartition, which allows us to introduce an effective temperature $T_\eff(t) = \e_k(t)\, n_k(t)$. This emergence of a temperature will be explained in the next section.

The excess energy in the large-momentum tail of the distribution may lead to a Kolmogorov-like cascade if the coupling between Bogoliubov modes is taken into account \cite{Buchhold2015}. This behavior is expected be present in 2D and 3D, leading to a true thermalization. However in 1D the Bogoliubov theory is believed to hold much better, so our dissipative state becomes a realization of a generalized Gibbs ensemble, where each mode is in a Gaussian state, but may have its own temperature \cite{Rigol2007,Langen2014}.

\section{Phononic limit}\label{sec:5}

\begin{figure}[t]
	\centering
	\includegraphics[width=0.9\linewidth]{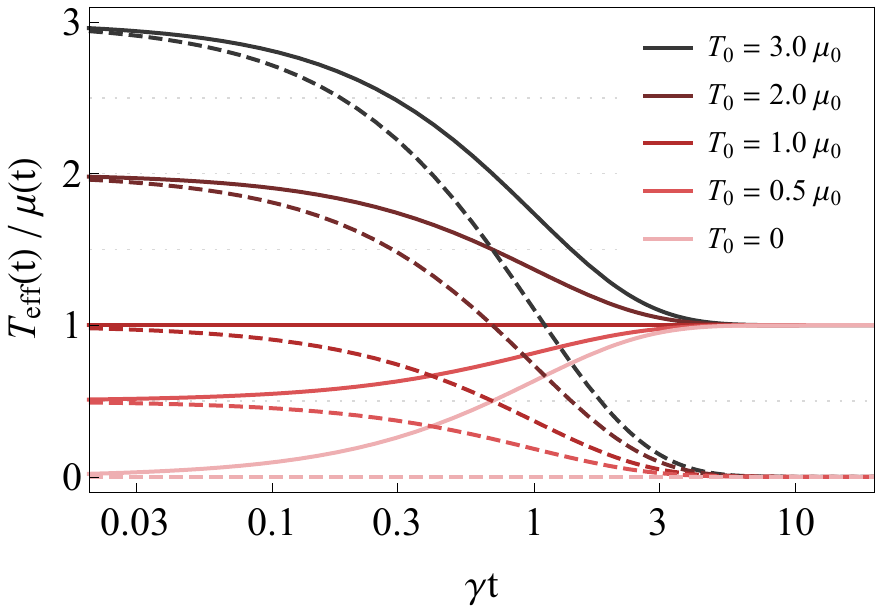}
	\caption{Temperature to the chemical potential ratio as a function of dimensionless time $\gamma t$ for different initial values of $T_0/\mu_0$ (solid lines). Note the emergence of an asymptotic dissipative state $T(t)= \mu(t)$ for $t \rightarrow \infty$. For comparison, we present the classical mean-field prediction which neglects the quantum noise (dashed lines). The scale of the time axis is logarithmic.}
	\label{fig:4}
\end{figure}

Emergence of the effective temperature $T_\eff$ can be proven in the low-energy phononic limit, which is recovered by considering the phase and density fluctuations on  length scales much larger than the condensate healing length $\xi(t) = 1/mc(t)$, where  $c(t) = \sqrt{\mu(t)/m}$ is a time-dependent speed of sound. Taking the phononic limit corresponds to neglecting the curvature of the dispersion relation
$\e_k(t) = c(t)\,|k| + O(k^2)$, and the Bogoliubov coefficients become
\begin{align}
u_k^2(t) = v_k^2(t) = u_k(t)\, v_k(t) = \dfrac{m c(t)}{2|k|} + O(k).
\label{ukvk}
\end{align}

Taking into account that the speed of sound decays exponentially $c(t) = c_0\, e^{-\gamma t}$, we can replace the time-dependent Bogoliubov coefficients with their initial values, e.g. $u_k(t) = u_k(0)\, e^{-\gamma t/2}$, then substitute \eqref{ukvk} into \eqref{nm2}, perform the integration and neglect terms of order $k$ and higher. This leads to 

\begin{align}
n_k(t) &= e^{-2\gamma t} n_k(0) + \tilde n_k(t),
\label{nm31}
\\
m_k(t) &= e^{-2(i\e_k+\gamma) t} m_k(0) + \tilde m_k(t),
\label{nm32}
\end{align}
where the quantum noise contributions are
\begin{align}
\tilde n_k(t) &= \frac{mc_0}{|k|}\, (e^{-\gamma t} - e^{-2\gamma t}),
 \\
  \tilde m_k(t) &= \frac{mc_0}{|k|} \dfrac{ (e^{-\gamma t} - e^{-2(i |k|c(t) + \gamma)t})}{1+2i\frac{|k|c(t)}{\gamma}}.
  \label{nm32q}
\end{align}

Expanding \eqref{nm32q} in the small parameter $\gamma$, we see that the quantum noise terms scale as $\left|\frac{\tilde m_k(t)}{\tilde n_k(t)}\right| \sim \frac{\gamma}{|k|c(t)}$, so the anomalous correlator contribution can be neglected in the experimentally relevant reigme of slow adiabatic dissipation $\gamma \ll c k$. This means that the system during dissipation is fully described by the modes' occupation numbers $n_k(t)$.

Defining an effective temperature in the phononic regime through classical equipartition $T_\eff = |k| c n_k$, from \eqref{nm31} we get
\begin{gather}
\label{Tmu}
\dfrac{T_\eff(t)}{\mu(t)} = \dfrac{T_0}{\mu_0} e^{-\gamma t} + \pa{1-e^{-\gamma t}},
\end{gather}
where 
the first term on the right-hand side comes from the mean-field, and the second term represents the contribution of the quantum noise. The initial state is assumed to be a true thermal equilibrium \eqref{initial} at temperature $T_0$. In a special case when the initial temperature is equal to the initial chemical potential, the two remain equal during the subsequent evolution (inset in Figure~\ref{fig:3}).

The effect of dissipation at the mean-field level can thus be understood as a removal of phonons from the system, which in the Rayleigh-Jeans approximation is directly equivalent to cooling. The quantum noise, on the other hand, creates new quasiparticles and may lead to re-heating. The competition between these two trends leads the evolution of the system
 towards an asymptotic state $T(t) = \mu(t)$, setting a limit on how far the system can be cooled through uniform Markovian dissipation (Figure~\ref{fig:4}). At long times semiclassical theory with Markovian quantum noise strongly deviates from the mean-field solution \eqref{UT.10mf},
 which would predict $\frac{T_\eff(t)}{\mu(t)} = \frac{T_0}{\mu_0} e^{-\gamma t}$ and hence $T_\eff/\mu \rightarrow 0$ (dashed lines in Figure~\ref{fig:4}).

As a convenient experimental probe, we propose to measure the temperature dependence on the chemical potential $\mu(t)$, which in the uniform case reads
\begin{gather}
\dfrac{T(t)}{T_0} = \pfrac{\mu(t)}{\mu_0}^{3/2}+\dfrac{\mu_0}{T_0} \bra{
	\dfrac{\mu(t)}{\mu_0} - \pfrac{\mu(t)}{\mu_0}^{3/2}}.
	\label{t1}
\end{gather}

\begin{figure}[t]
	\centering
	\includegraphics[width=0.8\linewidth]{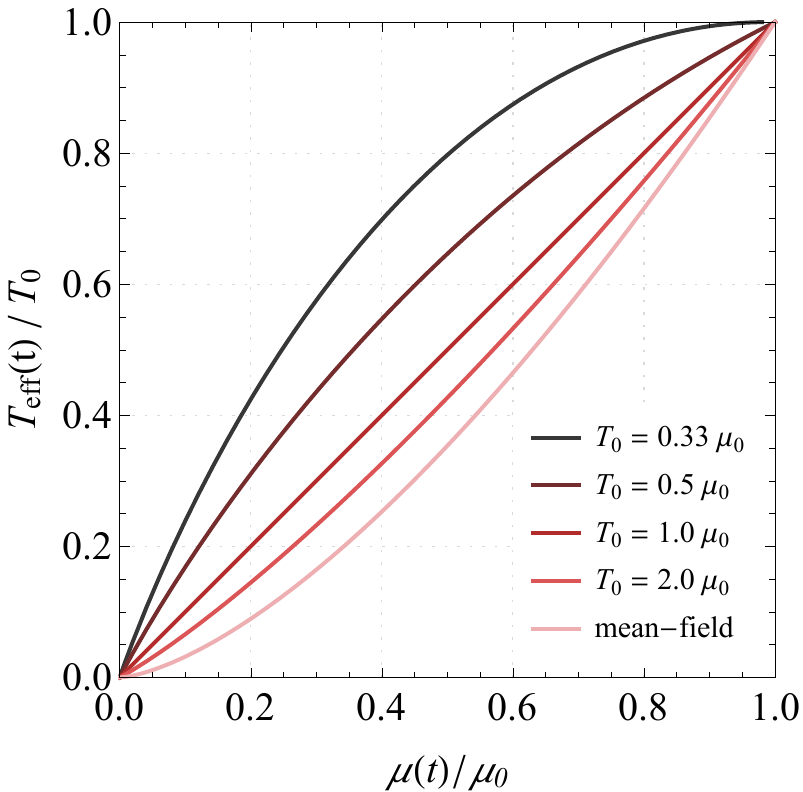}
	\caption{Effective temperature scaling with the time-dependent chemical potential for different $T_0/\mu_0$ ratios. The lowermost curve represents the classical limit where the quantum noise is neglected. The chemical potential is proportional to the mean density in the uniform case, and to the central peak density in case of the harmonic potential.}
	
	\label{fig:5}
\end{figure}

To test the predictions of \eqref{Tmu} and \eqref{t1} we propose the following experiment:
Prepare the quantum gas in a thermal state with known temperature and particle number. Then, turn on the rf or mw outcoupling, wait for an unspecified time $t$ and measure the number $N(t)$ and temperature $T(t)$ of the remaining particles.
 The temperature can be measured by switching off the confining trap and analyzing the density ripple pattern emerging in time-of-flight \cite{Imambekov2009,Manz2010}.
The chemical potential $\mu(t)$ can be calculated from the direct measurement of the total particle number $N(t)$, as in the uniform case $N(t) \propto \mu(t)$, and in the harmonically trapped case in the local density and Thomas-Fermi approximations $N(t) \propto \mu^{3/2}(t)$.
Predictions of \eqref{t1} are plotted in Figure~\ref{fig:5} for different initial ratios $T_0/\mu_0$, along with the mean-field prediction (mean-field and semiclassical results agree in the limit $T_0/\mu_0 \rightarrow \infty$). 

Eq.~\eqref{Tmu} can also be used to derive other important degeneracy criteria such as the scaling of the thermal coherence length $\lambda = \frac{2\mu}{m g T}$ and the Penrose-Onsager mode occupation number  $N_{PO} = \lambda N/L$. Namely, the gas becomes more degenerate during dissipation, as witnessed by the increasing coherence length $\lambda$ and the growing relative occupation of the Penrose-Onsager mode $N_{PO}/N$, as long as the initial temperature to chemical potential ratio $T_0/\mu_0>1$.

\section{Conclusions}
We developed a general theoretical description of dissipative degenerate Bose gases, where uniform Markovian dissipation is realized 
by outcoupling atoms from the condensate. Our model is applicable both for true condensates and quasicondensates at low temperatures as long as the Bogoliubov theory remains valid and conventional thermalization is suppressed. 

In one spatial dimension, we found that during dissipation the low-momentum phononic modes remain close to the thermal equilibrium, and that at a high enough initial temperature $T_0>\mu_0$ dissipation leads to cooling.
Due to the presence of a white quantum noise,  which stems from the
Markovian outcoupling to a continuum of  empty modes,
the systems evolves towards an asymptotic state with an effective temperature $T_\eff(t) = \mu(t)$ as $t\rightarrow\infty$.
 In addition, we presented scaling laws
for temperature dynamics, which can be used as guidelines for  experimental realizations.

 In higher dimensions, direct observations of the predicted effects may be limited mainly due to two reasons: firstly, the dissipative cooling may be overshadowed by the conventional evaporative cooling due to effective thermalization; and secondly, the outcoupling may be non-Markovian, e.g. as a result of the finite particle escape time.

A recent experiment with dissipative 1D condensates \cite{Rauer2015} measured the temperature dependence on the atom number and showed a much better agreement with the mean-field theory than the semiclassical one. We conjecture that it may be accounted for by the non-Markovianity of the outcoupling process, which can decrease the influence of the quantum noise (see discussion in Section~\ref{sec:noise}). The issue of non-Markovianity will be addressed in a following publication.

\acknowledgments
We are thankful to A.~Pol\-kov\-ni\-kov for valuable input at the beginning of this project and I.~Ca\-ru\-sotto for enlightening discussions and critical reading of the manuscript. We acknowledge support of the Austrian science fund (FWF) grant P22590--N16, ERC grant QuantumRelax, CoQuS doctoral program (PG and BR), and the Alexander von Humboldt Foundation through a Feodor Lynen Research Fellowship (TL).

\bibliography{library}

\end{document}